\documentclass[aps,prb,a4paper,preprint,showpacs]{revtex4}
\usepackage{amsmath}
\usepackage{amsfonts}
\usepackage{amssymb}
\usepackage{graphicx}
\usepackage{amsbsy}

\begin{document}
\title{Skyrmion dynamics in a chiral magnet driven by periodically varying spin currents }
\author{Rui Zhu\renewcommand{\thefootnote}{*}\footnote{Corresponding author.
Electronic address:
rzhu@scut.edu.cn} and Yin-Yan Zhang}
\address{Department of Physics, South China University of Technology,
Guangzhou 510641, People's Republic of China }

\begin{abstract}

In this work, we investigated the spin dynamics in a slab of chiral magnets induced by an alternating (ac) spin current. Periodic trajectories of the skyrmion in real space are discovered under the ac current as a result of the Magnus and viscous forces, which originate from the Gilbert damping, the spin transfer torque, and the $ \beta $-nonadiabatic torque effects. The results are obtained by numerically solving the Landau-Lifshitz-Gilbert equation and can be explained by the Thiele equation characterizing the skyrmion core motion.

\end{abstract}


\maketitle

\narrowtext

\section{Introduction}

The skyrmion spin texture is a kind of topologically-nontrivial magnetic vortex formed most typically in the bulk chiral magnets (CMs) and magnetic thin films\cite{NagaosaNatNano2013, MuhlbauerScience2009, HeinzeNatPhys2011}. In CMs it is believed that the spin-orbit coupling induced Dzyaloshinskii-Moriya interaction (DMI) governs the spin twisting\cite{NagaosaNatNano2013}. Recently the magnetic skyrmion structure attracts intensive focus, both in the fundamental theoretic aspect and in its potential application in the information technology\cite{XZhangSR2015, XXingPRB2016, YZhouNatCommun2014, SampaioNatNano2013}. In the magnetic skyrmion state the emergent electrodynamic effect originates from its nontrivial spin topology and gives rise to the topological Hall effect and a remarkable current-driven spin transfer torque effect\cite{NagaosaNatNano2013, HamamotoPRB2015, NeubauerPRL2009, SchulzNatPhys2012, RommingScience2013, TchoePRB2012, IwasakiNatNano2013, RalphJMMM2008}. The so-called skyrmionics makes use of the skyrmion as a memory unit favored by its topologically protected long lifetime and ultralow driving current, which is five or six orders smaller than that for driving a magnetic domain wall\cite{NagaosaNatNano2013, JonietzScience2010}.

Although the current-driven spin dynamics in the CMs with DMI has been intensively studied recently, less work of an alternating current (ac) driven skyrmion dynamics was reported. The skyrmion-motion-induced ac current generation has been predicted, which shares the reversed effect of our consideration\cite{SZLinPRL2014}. In this work, we investigated the ac-spin-current driven skyrmion dynamics with the DMI, Gilbert damping\cite{HalsPRB2014}, adiabatic and nonadiabatic spin torques, and different current profiles taken into account. Our proposition is inspired by the following several aspects. Firstly, it is both theoretically and technically interesting to know the behavior of a skyrmion when an external ac current is applied. Secondly, a high-speed low-power modulation of a skyrmion is favorable for potential memory processing. Lastly but not least, we noticed the mathematical significance of the solution of the Landau-Lifshitz-Gilbert (LLG) equation of a collinear magnet with periodically varying spin-currents applied, in which chaos is observed\cite{LakshmananPTRSA2011, ZYangPRL2007}.

The topological property of a spin texture can be described by the surface integral of the solid angle of the unitary spin-field vector ${\bf{n}}({\bf{r}})$. The skyrmion number is so defined as $S = \frac{1}{{4\pi }}\int {{\bf{n}} \cdot \left( {\frac{{\partial {\bf{n}}}}{{\partial x}} \times \frac{{\partial {\bf{n}}}}{{\partial y}}} \right){d^2}{\bf{r}}} $ counting how many times the spin field wraps the unit sphere. More specific topological properties of a skyrmion can be considered by analyzing its radial and whirling symmetric pattern\cite{NagaosaNatNano2013, XZhangPRB2016, BogdanovJMMM1999, YYZhangPLA2016}. In the continuum field theory, as a result of topological protection, the skyrmion cannot be generated from a topologically trivial magnetic state such as a ferromagnet or a helimagnet by variation without a topologically nontrivial force such as a spatially nontrivial spin current\cite{RommingScience2013, TchoePRB2012}, geometrical constriction\cite{IwasakiNatNano2013}, domain wall pair source\cite{YZhouNatCommun2014, XXingPRB2016}, the edge spin configuration\cite{YYZhangPLA2016}, etc., and vise versa. It is predicted by simulation that the skyrmion can be generated from a quasi-ferromagnetic and helimagnetic state by external Lorentzian and radial spin current\cite{TchoePRB2012} and that transformation is possible between different topologically-nontrivial states such as that between the domain-wall pair and the skyrmion\cite{YZhouNatCommun2014, XXingPRB2016}. The local current flowing from the scanning tunneling microscope to generate the skyrmion in experiment can be approximated by a radial spin current, which imbues nontrivial topology into the helimagnet\cite{RommingScience2013}. Also an artificial magnetic skyrmion can be tailored by an external magnetic field with nontrivial geometric distribution\cite{JLiNatCommun2014}. When the boundary geometry of the material is tailored such as by a notch in a long plate, a skyrmion can be generated by a collinear spin current\cite{IwasakiNatNano2013}. In this case, the nontrivial constriction topology contributes to the formation of the skyrmion. The uniform current can move and rotate a skyrmion without changing its topology\cite{JonietzScience2010, EverschorPRB2012}. In this work we will show that these topological behaviors of the skyrmion are retained in the spin dynamics driving by an ac spin current.

Almost all kinds of ferromagnetic and vortex spin dynamics can be described by the LLG equation. The behavior of the LLG equation is of importance in both the physical and mathematical sciences\cite{LakshmananPTRSA2011}. It has been shown by previous works that the spin torque effect driven by a periodic varying spin current can be described as well by the LLG equation with the original time-independent current replaced by the time-dependent current in the spin torque term\cite{LakshmananPTRSA2011, ZYangPRL2007}. Although chaotic behaviors are predicted in the spatially-uniform ac spin-current driven collinear ferromagnetic spin structure\cite{LakshmananPTRSA2011, ZYangPRL2007}, which is well described by the single-spin LLG equation, no similar phenomenon is reported in a spatially-nonuniform spin lattice, the latter of which can be attributed to the relaxation processes of the inter-site scattering. Even if some sort of chaotic behavior occurs after a long time of evolution, it is workable to restore the original state by applying a magnetic field after some time. The influence of it on the skyrmionics exploitation is not large. In this work, we use a matrix-based fourth-order Runge-Kutta method to solve the LLG equation with both the adiabatic and nonadiabatic spin torques taken into account. Analytical solution of the generalized Thiele equation\cite{NagaosaNatNano2013, EverschorPRB2012, IwasakiNatNano2013} reproduces our numerical results.

\section{Theoretic Formalism }

 We consider a thin slab of CM modulated by a constant magnetic field and an ac spin current. The strong DMI makes the material a skyrmion-host. In the continuum approximation, the Hamiltonian of the localized magnetic spin in a CM can be described as\cite{NagaosaNatNano2013, IwasakiNatNano2013, TchoePRB2012}
\begin{equation}
\begin{array}{l}
H =  - J\sum\limits_{\bf{r}} {{{\bf{M}}_{\bf{r}}} \cdot \left( {{{\bf{M}}_{{\bf{r}} + {{\bf{e}}_x}}} + {{\bf{M}}_{{\bf{r}} + {{\bf{e}}_y}}}} \right)} \\
 - D\sum\limits_{\bf{r}} {\left( {{{\bf{M}}_{\bf{r}}} \times {{\bf{M}}_{{\bf{r}} + {{\bf{e}}_x}}} \cdot {{\bf{e}}_x} + {{\bf{M}}_{\bf{r}}} \times {{\bf{M}}_{{\bf{r}} + {{\bf{e}}_y}}} \cdot {{\bf{e}}_y}} \right)} \\
 - {\bf{B}} \cdot \sum\limits_{\bf{r}} {{{\bf{M}}_{\bf{r}}}} ,
\end{array}
\end{equation}
with $J$ and $D$ the ferromagnetic and Dzyaloshinskii-Moriya (DM) exchange energies, respectively. The dimensionless local magnetic moments ${\bf {M}}_{\bf {r}}$ are defined as ${{\bf{M}}_{\bf{r}}} \equiv  - {{{{\bf{S}}_{\bf{r}}}} \mathord{\left/
 {\vphantom {{{{\bf{S}}_{\bf{r}}}} \hbar }} \right.
 \kern-\nulldelimiterspace} \hbar }$, where ${\bf {S}}_{\bf {r}}$ is the local spin at ${\bf {r}}$ and $\hbar$ is the plank constant divided by $2 \pi$. We assume that the length of the vector $\left| {{{\bf{M}}_{\bf{r}}}} \right| = M$ is fixed, therefore ${{\bf{M}}_{\bf{r}}} = M{\bf{n}}\left( {\bf{r}} \right)$ with ${\bf{n}}\left( {\bf{r}} \right)$ the unitary spin field vector. The unit-cell dimension is taken to be unity. An external magnetic field $\bf{B}$ is applied perpendicular to the slab plane to stabilize the skyrmion configuration. The Bohr magneton $\mu _B$ is absorbed into $\bf{B}$ to have it in the unit of energy. The typical DMI $D=0.18J$ is used throughout this work\cite{IwasakiNatNano2013}. This DM exchange strength corresponds to the critical magnetic fields $B_{c1}=0.0075J$ between the helical and skyrmion-crystal phases and $B_{c2}=0.0252J$ between the skyrmion-crystal and ferromagnetic phases, respectively. We adopt ${\bf{B}}=(0,0,0.01J)$ in our numerical considerations with $J=1$ meV.

The extended form of the LLG equation that takes into account the DMI and the adiabatic and nonadiabatic spin torque effects can be expressed in the following formula\cite{NagaosaNatNano2013, MironNatMater2010, IwasakiNatNano2013, TchoePRB2012}
\begin{equation}
\begin{array}{l}
\frac{{d{{\bf{M}}_{\bf{r}}}}}{{dt}} =  - \gamma {{\bf{M}}_{\bf{r}}} \times {\bf{B}}_{\bf{r}}^{{\rm{eff}}} + \frac{\alpha }{M}{{\bf{M}}_{\bf{r}}} \times \frac{{d{{\bf{M}}_{\bf{r}}}}}{{dt}} + \frac{{p{a^3}}}{{2eM}}\left[ {{\bf{j}}\left( {{\bf{r}},t} \right) \cdot \nabla } \right]{{\bf{M}}_{\bf{r}}}\\
 - \frac{{p{a^3}\beta }}{{2e{M^2}}}\left\{ {{{\bf{M}}_{\bf{r}}} \times \left[ {{\bf{j}}\left( {{\bf{r}},t} \right) \cdot \nabla } \right]{{\bf{M}}_{\bf{r}}}} \right\}.
\end{array}
\label{LLG Equation}
\end{equation}
By assuming that the energy of a magnet with the local magnetization ${\bf {M}}_{\bf {r}}$ in a spatially varying magnetic field ${{\bf{B}}_{\bf{r}}^{{\rm{eff}}}}$ has the form of $H =  - \gamma \hbar \sum\limits_{\bf{r}} {{{\bf{M}}_{\bf{r}}} \cdot {\bf{B}}_{\bf{r}}^{{\rm{eff}}}} $, we have
\begin{equation}
{\bf{B}}_{\bf{r}}^{{\rm{eff}}} =  - \frac{1}{{\hbar \gamma }}\frac{{\partial H}}{{\partial {{\bf{M}}_{\bf{r}}}}},
\end{equation}
and therefore the first term in the right hand side of Eq. (\ref{LLG Equation}) is
\begin{equation}
\begin{array}{l}
 - \gamma {{\bf{M}}_{\bf{r}}} \times {\bf{B}}_{\bf{r}}^{{\rm{eff}}} =  - \frac{J}{\hbar }{{\bf{M}}_{\bf{r}}} \times \left( {{{\bf{M}}_{{\bf{r}} + {{\bf{e}}_x}}} + {{\bf{M}}_{{\bf{r}} + {{\bf{e}}_y}}}{\rm{ + }}{{\bf{M}}_{{\bf{r}}{\rm{ - }}{{\bf{e}}_x}}} + {{\bf{M}}_{{\bf{r}}{\rm{ - }}{{\bf{e}}_y}}}} \right) - \frac{1}{\hbar }\left( {{{\bf{M}}_{\bf{r}}} \times {\bf{B}}} \right)\\
 - \frac{D}{\hbar }{{\bf{M}}_{\bf{r}}} \times \left[ \begin{array}{l}
\left( {{{\bf{M}}_{{\bf{r}} - {{\bf{e}}_y},z}} - {{\bf{M}}_{{\bf{r}} + {{\bf{e}}_y},z}}} \right){{\bf{e}}_x} + \left( {{{\bf{M}}_{{\bf{r}} + {{\bf{e}}_x},z}} - {{\bf{M}}_{{\bf{r}} - {{\bf{e}}_x},z}}} \right){{\bf{e}}_y}\\
 + \left( {{{\bf{M}}_{{\bf{r}} + {{\bf{e}}_y},x}} - {{\bf{M}}_{{\bf{r}} + {{\bf{e}}_x},y}} - {{\bf{M}}_{{\bf{r}} - {{\bf{e}}_y},x}} + {{\bf{M}}_{{\bf{r}} - {{\bf{e}}_x},y}}} \right){{\bf{e}}_z}
\end{array} \right].
\end{array}
\label{Effective Magnetic Field}
\end{equation}
The second to the last terms of Eq. (\ref{LLG Equation}) sequentially correspond to the effect of the Gilbert damping, the time-dependent spin current ${\bf {j}} (t)={\bf {j}}_e \sin (\omega t)$-induced adiabatic and nonadiabatic spin torques, respectively. $p$ measures the polarization of the conduction electrons, $e$ is the positive electron charge, and $a$ is the average in-plane lattice constant of the CM. In our considerations the frequency of the ac spin current $\omega $ is small enough in comparison of the magnetization evolution rate. Therefore, the spin torques can be satisfactorily described by using the time-dependent current in the standard torque expression, which has been justified by previous studies\cite{LakshmananPTRSA2011, ZYangPRL2007}. Here, the unit of time is set to be $t_0= \hbar /J \approx 6.6 \times 10^{-13}$ s. A phenomenologically expected value of $\alpha = 0.1$ is used in afterwards numerical considerations.

By looking deep into Eqs. (\ref{LLG Equation}) and (\ref{Effective Magnetic Field}), we can make some predictions of the behavior of the local magnetization. We know that the effect of the magnetic field together with the Gilbert term is to precess the magnetic spin into the direction of the external field. The first term in the right hand side of Eq. (\ref{Effective Magnetic Field}) is that the effective magnetic field is in the direction of neighboring spins. Therefore the evolution tends to form a ferromagnet. This contributes to the centripetal force of the magnetization in the direction of ${{\bf{M}}_{\bf{r}}} \times {{\bf{M}}_{{\bf{r}}'}}$, which results in the precession of one around the other. The effective field in the DM term in Eq. (\ref{Effective Magnetic Field}) is $ - \nabla  \times {{\bf{M}}_{\bf{r}}}$ with unitary lattice constants. The integral counterpart of the curl is $\oint {{{\bf{M}}_{{\bf{r}}'}} \cdot d{\bf{l}}} $. When the neighboring spins form a ring, the energy is the lowest, hence generating a spiraling force to the CM. It helps our understanding if we analogize all the other terms in the right hand side of Eq. (\ref{LLG Equation}) to the effect of a magnetic field. The local ``magnetic field" of the phenomenological Gilbert damping force is proportional to $ - {{d{{\bf{M}}_{\bf{r}}}} \mathord{\left/
 {\vphantom {{d{{\bf{M}}_{\bf{r}}}} {dt}}} \right.
 \kern-\nulldelimiterspace} {dt}}$ in the standard linear-response damping form, proportional to the velocity and pointing oppositely to it. While the local spin is precessing, the direction of ${{\bf{M}}_{\bf{r}}} \times {{d{{\bf{M}}_{\bf{r}}}} \mathord{\left/
 {\vphantom {{d{{\bf{M}}_{\bf{r}}}} {dt}}} \right.
 \kern-\nulldelimiterspace} {dt}}$ points to the precession axis of ${{\bf{M}}_{\bf{r}}}$ adding a force swaying to that axis. The last two terms are the effect of the current-induced spin torques. For convenience of interpretation, we discuss the case that ${{\bf{j}}\left( {{\bf{r}},t} \right)}$ is spatially-uniform and along the $x$-direction. Then $\left[ {{\bf{j}}\left( {{\bf{r}},t} \right) \cdot \nabla } \right]{{\bf{M}}_{\bf{r}}} = {j_x}\left( t \right){{\partial {{\bf{M}}_{\bf{r}}}} \mathord{\left/
 {\vphantom {{\partial {{\bf{M}}_{\bf{r}}}} {\partial x}}} \right.
 \kern-\nulldelimiterspace} {\partial x}}$. In the case of the adiabatic torque, this term adds a velocity to ${{{\bf{M}}_{\bf{r}}}}$ making it sway to the direction of ${{{\bf{M}}_{{\bf{r}} + {{\bf{e}}_x}}}}$, and ${{{\bf{M}}_{{\bf{r}} + {{\bf{e}}_x}}}}$ to ${{\bf{M}}_{{\bf{r}} + 2{{\bf{e}}_x}}}$, and etc. if ${j_x}\left( t \right)$ is positive. Therefore, the complete spin texture moves along the direction of the external spin current like a relay race no matter it is a skyrmion or a domain wall. In a periodic magnetic structure such as a ferromagnet and a helimagnet the ``relay race" goes back to itself and hence no spin structure movement occurs. Following this physical picture, the local ``magnetic field" of the nonadiabatic spin torque is along the direction of ${j_x}\left( t \right){{\partial {{\bf{M}}_{\bf{r}}}} \mathord{\left/
 {\vphantom {{\partial {{\bf{M}}_{\bf{r}}}} {\partial x}}} \right.
 \kern-\nulldelimiterspace} {\partial x}}$. It exerts a velocity perpendicular to that originates from the adiabatic spin torque. Its result is the motion of the spin texture in the direction perpendicular to the spin current. We have already analyzed the mechanisms of the LLG equation term by term. However, they affects the system collaboratively. While the adiabatic spin torque moves the skyrmion along the spin current, the Gilbert damping force contributes a velocity in the direction of ${{\bf{M}}_{\bf{r}}} \times {{d{{\bf{M}}_{\bf{r}}}} \mathord{\left/
 {\vphantom {{d{{\bf{M}}_{\bf{r}}}} {dt}}} \right.
 \kern-\nulldelimiterspace} {dt}}$ and therefore the effect is a transverse motion of the skyrmion, which is the so-called Hall-like motion\cite{TchoePRB2012}. Also it is noticeable that the transverse velocity resulting from the Gilbert damping and the nonadiabatic spin torque is opposite to each other. In real situations, both $\alpha $ and $\beta $ are much less than 1. The adiabatic spin torque makes the main contribution to the motion of the skyrmion. And when the two transverse force is equal, the motion of the skyrmion is straightly along the direction of the spin current. Therefore, periodic trajectories of the skyrmion in real space can be predicted under the influence of a spatially uniform ac spin current.

The previous discussions are well expressed in the Thiele equation describing the motion of the center of mass of a skyrmion as\cite{NagaosaNatNano2013, EverschorPRB2012, ThielePRL1972, IwasakiNatNano2013}.
\begin{equation}
{\bf{G}} \times \left[ { - {\bf{j}}\left( t \right) - {{\bf{v}}_d}} \right] + \kappa \left[ { - \beta {\bf{j}}\left( t \right) - \alpha {{\bf{v}}_d}} \right] -\nabla U (\bf {r})= 0,
\label{Thiele Equation}
\end{equation}
where ${{\bf{v}}_d} = {{d{\bf{R}}} \mathord{\left/
 {\vphantom {{d{\bf{R}}} {dt}}} \right.
 \kern-\nulldelimiterspace} {dt}} = \left( {\dot X,\dot Y} \right)$ with ${\bf{R}} = \left( {X,Y} \right)$ the center of mass coordinates, $\kappa $ is a dimensionless constant of the order of unity, and ${\bf{G}} = 2\pi S{{\bf{e}}_z}$ is the gyrovector with ${\bf {e}}_z$ in the direction perpendicular to the CM plane. The minus sign before ${\bf {j}} (t)$ is because of that the direction of the motion of conduction electrons is opposite to that of the current. The Thiele equation (\ref{Thiele Equation}) describes\cite{XXingPRB2016} coaction of the Magnus force ${{\bf{F}}_{\rm{g}}} = {\bf{G}} \times \left[ { - {\bf{j}}\left( t \right) - {{\bf{v}}_d}} \right]$, the viscous force ${{\bf{F}}_{\rm{v}}} = \kappa \left[ { - \beta {\bf{j}}\left( t \right) - \alpha {{\bf{v}}_d}} \right]$, and the confining force ${{\bf{F}}_{\rm{p}}} =  - \nabla U\left( {\bf{r}} \right)$. In our considerations, periodic boundary conditions are used to justify an infinite two-dimensional model. The applied magnetic field is spatially uniform and the impurity effect is neglected. Therefore $\nabla U \approx 0$. The analytical result of Eq. (\ref{Thiele Equation}) assuming $S=-1$ and ${\bf{j}}\left( t \right) = {j_e}\sin \left( {\omega t} \right){{\bf{e}}_x}$ can be obtained as
 \begin{equation}
\left\{ \begin{array}{l}
 X =   \frac{{\alpha \beta {\kappa ^2} + 4{\pi ^2}}}{{\left( {{\alpha ^2}{\kappa ^2} + 4{\pi ^2}} \right)\omega }}{j_e}\cos \left( {\omega t} \right),\\
 Y = \frac{{2\pi \kappa \left( {\beta  - \alpha } \right)}}{{\left( {{\alpha ^2}{\kappa ^2} + 4{\pi ^2}} \right)\omega }}{j_e}\cos \left( {\omega t} \right).
\end{array} \right.
\label{Cosinusoidal Trajectory}
\end{equation}
Since the spin current is time dependent, ${{\bf{F}}_g}$ and ${{\bf{F}}_{\rm{v}}}$ instantaneously change their direction with the motion of the skyrmion core and simultaneously react on the motion of the skyrmion, giving rise to the trigonometric trajectory of the skyrmion shown in Eq. (\ref{Cosinusoidal Trajectory}), which agrees with the simulation results. Because the skyrmion vortex moves in the relay fashion under the effect of the spin torque shown by the LLG equation, there is a $\pi /2$ phase lag between its core motion and the sinusoidally varying spin current.

\section{Numerical Results and Interpretations}

By multiplying ${\tilde \alpha ^{ - 1}}$ with
\begin{equation}
\tilde \alpha  = 1 - \alpha \left[ {\begin{array}{*{20}{c}}
0&{ - {{\left( {{{\bf{M}}_{\bf{r}}}} \right)}_z}}&{{{\left( {{{\bf{M}}_{\bf{r}}}} \right)}_y}}\\
{{{\left( {{{\bf{M}}_{\bf{r}}}} \right)}_z}}&0&{ - {{\left( {{{\bf{M}}_{\bf{r}}}} \right)}_x}}\\
{ - {{\left( {{{\bf{M}}_{\bf{r}}}} \right)}_y}}&{{{\left( {{{\bf{M}}_{\bf{r}}}} \right)}_x}}&0
\end{array}} \right],
\end{equation}
from the left to Eq. (\ref{LLG Equation}), the matrix-based Runge-Kutta method is developed. In Figs. 1 to 3, numerical results of our simulations are given. We set $M=1$, $p=0.5$, and $a=4$ $\rm {\AA}$. The integral step $h=0.1 t_0$ is used and its convergence is justified by comparison with the results of $h=0.01 t_0$. With $D=0.18J$, the natural helimagnet wavevector $Q = {{2\pi } \mathord{\left/
 {\vphantom {{2\pi } \lambda }} \right.
 \kern-\nulldelimiterspace} \lambda } = {D \mathord{\left/
 {\vphantom {D J}} \right.
 \kern-\nulldelimiterspace} J}$ with the diameter of the skyrmion $\lambda  = {D \mathord{\left/
 {\vphantom {D J}} \right.
 \kern-\nulldelimiterspace} J} \approx 35$ in the unit of $a$. A ${\rm{30}} \times {\rm{30}}$ square lattice is considered which approximately sustains a single skyrmion. Periodic boundary condition is used to allow the considered patch to fit into an infinite plane. While part of the skyrmion moves out of the slab, complementary part enters from the outside as the natural ground state of a CM is the skyrmion crystal. We use the theoretically perfect skyrmion profile ${\bf{n}}\left( {\bf{r}} \right) = \left[ {\cos \Phi \left( \varphi  \right)\sin \Theta \left( r \right),\sin \Phi \left( \varphi  \right)\sin \Theta \left( r \right),\cos \Theta \left( r \right)} \right]$ with $\Theta \left( r \right) = \pi \left( {1 - {r \mathord{\left/
 {\vphantom {r \lambda }} \right.
 \kern-\nulldelimiterspace} \lambda }} \right)$ and $\Phi \left( \varphi  \right) = \varphi $ in the polar coordinates as the initial state and it would change into a natural skyrmion in less than one current period. The skyrmion number for this state $S=-1$. The spatially-uniform ac spin current is applied in the $x$-direction as ${\bf{j}}\left( t \right) = {j_e}\sin \left( {\omega t} \right){{\bf{e}}_x}$ within the CM plane.

Variation of the skyrmion number in time driven by the ac spin current is shown in Fig. 1. It can be seen that cosinusoidal variation of $S$ originates from the sinusoidal ${\bf{j}}\left( t \right)$ with exactly the same period. Fig. 2 shows the snapshots of the spin profile at the bottoms and peeks of the cosinusoidal variation of $S$ and Fig. 3 shows the trajectories of the center of the skyrmion (see Ref. \onlinecite{SupplementaryMovie}
for Supplementary Movie). The skyrmion number is a demonstration of the motion pattern of the skyrmion. While the skyrmion moves to one side of the CM slab, only part of a skyrmion is within the view and hence the skyrmion number is reduced. Previous authors have found that the velocity of the skyrmion increases linearly with the increase of the current amplitude and that the dynamical threshold current to move a skyrmion is in the same order of that needed for a domain wall\cite{NagaosaNatNano2013}. Here we have reobtained the two points. It can be seen in Fig. 1(a) that the peak hight of the skyrmion number increases with the amplitude of the current density and it becomes almost invisible when $j_e$ is as small as $10^{10}$ A$\rm {m^{-2}}$. In Fig. 1(b), the evolutions of $S$ for different ac periods are shown. The frequency of the ac current is in the order of GHz, which is sufficiently adiabatic as the rate of the spin dynamics is in the order of $10^{-12}$ s. We can see that the periodic pattern of $S$ is better kept with larger amplitudes for smaller ac frequencies. It shows that the phenomenon is a good adiabatic one. Within our numerical capacity, it can be predicted that strong cosinusoidal variation can occur at MHz or smaller ac frequencies, which promises experimental realization.

The variation of $S$ is the result of the motion of the skyrmion. The periodic translation of skyrmion is the result of the coaction of the instantaneous Magnus and viscous forces. The spin current gives rise to the drift velocity of the spin texture. As a combined result of the Gilbert damping, the DMI, the adiabatic and nonadiabatic spin torques, the skyrmion Hall effect, namely, the transverse motion of the skyrmion perpendicular to the spin current, is observed in topologically-nontrivial spin textures. As shown in Fig. 2, in spite of its motion, the topological properties of the skyrmion are conserved because the initial skyrmion state and the natural skyrmion ground state share similar topology and no topology-breaking source such as an in-plane magnetic field is present. When part of the skyrmion moves out of the CM slab, only the remaining part contributes to the skyrmion number and hence $S$ is decreased. The cosinusoidal variation of $S$ directly reflects the oscillating trajectory of the skyrmion Shown in Figs. 2 and 3. We can see that the skyrmion changes from the initial artificial skyrmion state into the natural skyrmion state with $S=-1$ conserved, as shown in Fig. 2(a) and (b). At the times of integer periods the skyrmion is at the center of the CM slab and at the times of half-integer periods the skyrmion moves to the left side as shown in Fig. 2 (c) to (f).

As predicted by the Thiele equation, the trajectory of the skyrmion follows a cosinusoidal pattern expressed in Eq. (\ref{Cosinusoidal Trajectory}). It is interesting that the trajectory of the skyrmion results from the competition between the drift motion of the skyrmion and the skyrmion Hall effect under the influence of the adiabatic and nonadiabatic spin torque effects. The adiabatic spin torque effect exerts a force to align the spin at each site to its $+x$-direction neighbor while ${\bf {j}} (t)$ is in the ${\bf {e}} _x$ direction, which results in the motion of the spin pattern to the $+x$ direction in a relay fashion. The Gilbert damping effect and the nonadiabatic spin torque add a transverse velocity to the moving skyrmion perpendicular to its original velocity. These two forces are in opposite directions when $\alpha$ and $\beta$ are both positive. Therefore the transverse motion is determined by the sign and relative strength of these two effects. From Eq. (\ref{Cosinusoidal Trajectory}) we can see that when $\beta - \alpha  >0$ the skyrmion's $y$-direction motion is in a cosinusoidal form and when $\beta - \alpha  <0$ it is in a negative cosinusoidal form. For the $x$-direction motion of the skyrmion, the direction is the same in the two cases and the magnitude is slightly smaller for the latter because $\left| {4{\pi ^2}} \right| \gg \left| {\alpha \beta {\kappa ^2}} \right|$ holds for all physical parameter settings. And physically it is because the $x$-direction motion of the skyrmion is mainly determined by the adiabatic spin torque, which is the prerequisite for any motion of the skyrmion.

Our simulation results of the skyrmion trajectories for $\beta =0.5 \alpha$, $\alpha$, and $2 \alpha$ with fixed $\alpha =0.1$ are shown in Fig. 3. Good agreement with the prediction by the Thiele equation is obtained. In the three cases, $X$ evolves cosinusoidally with the initial position $(X,Y)=(15,15)$ at the center of the CM slab. For $\beta = 0.5 \alpha$, $Y$ evolves minus-cosinusoidally; for $\beta = \alpha$, $Y$ is constant at $15$; for $\beta = 2 \alpha$, $Y$ evolves cosinusoidally. As the difference between $\beta$ and $\alpha$ is small in Fig. 3 (a) and (c), the cosinusoidal pattern shrinks into a step jump. Besides the oscillation, a tiny linear velocity of the skyrmion can be seen in Fig. 3 (a) and (c). And the directions of this velocity are different in the two cases. We attribute this linear velocity to the whirling of the skyrmion from the artificial initial profile to the natural profile sustained by the real CM. Because at this whirling step, the Gilbert damping and the adiabatic and nonadiabatic torques are already in effect, the initial linear velocities are different in the two cases.

\section{Conclusions}

In this work, we have investigated the dynamics of the skyrmion in a CM driven by periodically varying spin currents by replacing the static current in the LLG equation by an adiabatic time-dependent current. Oscillating trajectories of the skyrmion are found by numerical simulations, which are in good agreement with the analytical solution of the Thiele equation. In the paper, physical behaviors of the general LLG equation with the Gilbert damping and the adiabatic and nonadiabatic spin torques coexistent are elucidated. Especially, the effect of the nonadiabatic spin torque is interpreted both physically and numerically.

\section{Author contribution statement}

R.Z. wrote the program and the paper. Y.Y.Z. made the simulation.

\section{Acknowledgements}

R.Z. would like to thank Pak Ming Hui for stimulation and encouragement of the work. This project was supported by the National Natural Science
Foundation of China (No. 11004063) and the Fundamental Research
Funds for the Central Universities, SCUT (No. 2014ZG0044).

\clearpage

\begin{figure}[h]
\includegraphics[height=10cm, width=14cm]{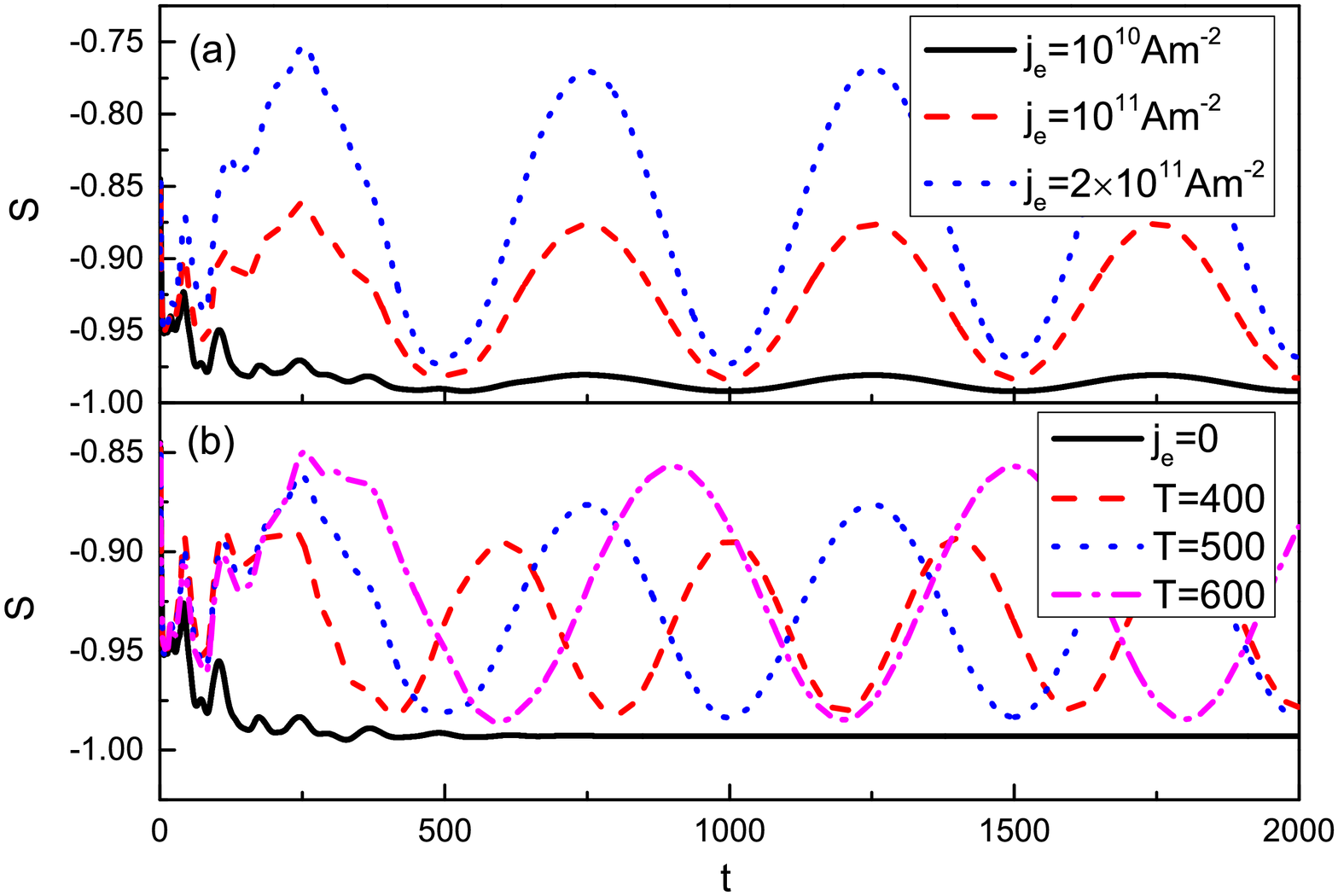}
\caption{Variation of the skyrmion number S in time (a) for different current amplitudes and (b) for different ac current frequencies. The time $t$ and ac spin current period $T$ are in the unit of $t_0 \approx 6.6 \times 10^{-13}$ s. $\beta =0.05$. In panel (a), $T=500$. In panel (b), $j_e =10^{11}$ A$\rm {m^{-2}}$.     }
\end{figure}

\clearpage

\begin{figure}[h]
\includegraphics[height=10cm, width=17cm]{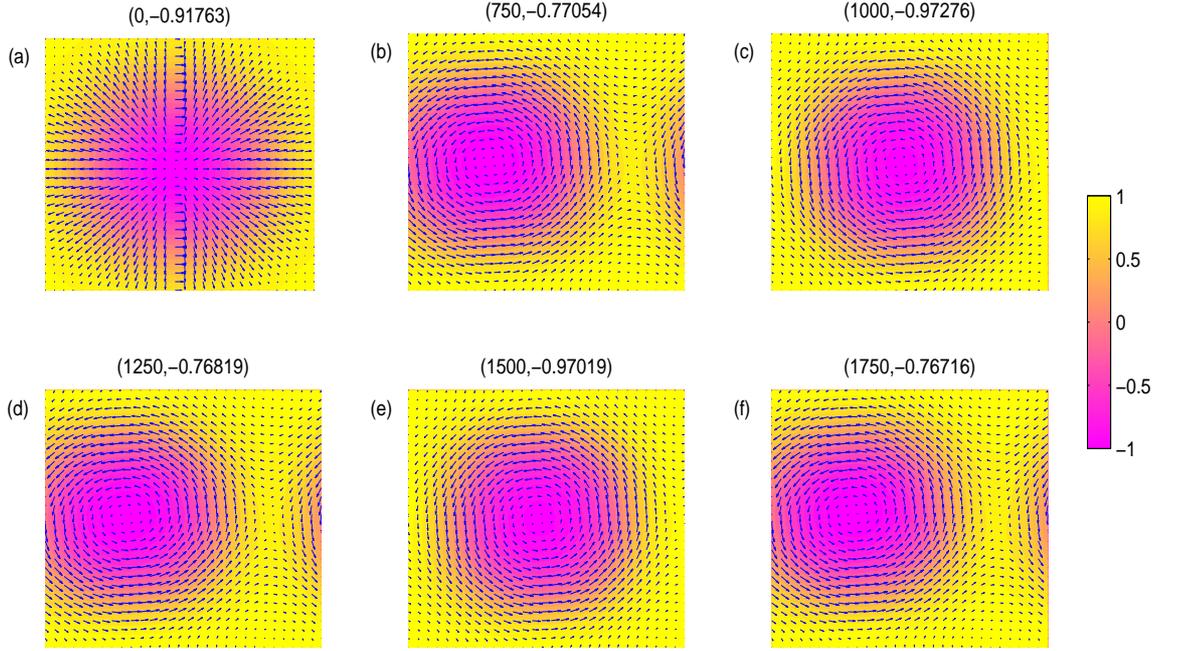}
\caption{ Snapshots of the dynamical spin configurations at the bottoms and peaks of the skyrmion number shown in Fig. 1. The in-plane components of the magnetic moments are represented by arrows and their $z$-components are represented by the color plot. The parameters are $j_e =2 \times 10^{11}$ A$\rm {m^{-2}}$, $T=500$, and $\beta =0.05$. On the top of each panel are the $(t,S)$ values. }
\end{figure}

\clearpage

\begin{figure}[h]
\includegraphics[height=10cm, width=14cm]{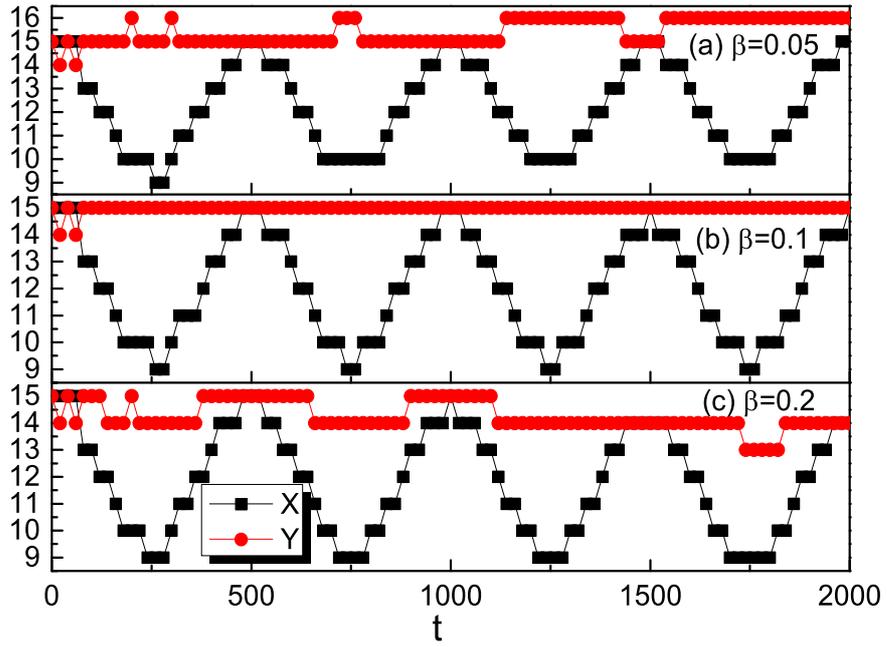}
\caption{Variation of the skyrmion center coordinates $(X,Y)$ in time (a) for $\beta =0.05$, (b) for $\beta =0.1$, and (c) for $\beta =0.2$. Other parameters are the same as Fig. 2.   }
\end{figure}

\end{document}